\documentclass[numbers]{elsarticle}
\usepackage{latexsym}
\usepackage{float}
\usepackage[latin2]{inputenc}
\usepackage{graphicx}
\usepackage{ae}
\usepackage{aecompl}
\usepackage{setspace}
\usepackage{booktabs}
\usepackage{longtable}
\usepackage{mciteplus}
\newtheorem{theorem}{Theorem}

\newtheorem{definition}{Definition}

\newtheorem{example}[theorem]{Example}

\usepackage[hidelinks]{hyperref}

\makeatletter
\def\ps@pprintTitle{%
	\let\@oddhead\@empty
	\let\@evenhead\@empty
	\def\@oddfoot{\centerline{\thepage}}%
	\let\@evenfoot\@oddfoot}
\renewcommand\@biblabel[1]{(#1)} 
\makeatother

\title{Navigating Homogeneous Paths through Amyloidogenic and Non-Amyloidogenic Hexapeptides}
		
		\author[p]{László Keresztes\corref{cor2}}
		\ead{keresztes@pitgroup.org}
		\author[p]{Evelin Szögi\corref{cor2}}
		\ead{szogi@pitgroup.org}
		\author[p]{Bálint Varga}
		\ead{balorkany@pitgroup.org}
		\author[s]{Viktor Farkas}
		\ead{farkasv@caesar.elte.hu}
		\author[s,t]{András Perczel}
		\ead{perczel@chem.elte.hu}
		\author[p,u]{Vince Grolmusz\corref{cor1}}
		\ead{grolmusz@pitgroup.org}
		\cortext[cor1]{Corresponding author}
		\cortext[cor2]{Joint first authors}
		\address[p]{PIT Bioinformatics Group, Eötvös University, H-1117 Budapest, Hungary}
		\address[u]{Uratim Ltd., H-1118 Budapest, Hungary}
		\address[s]{ELKH-ELTE Protein Modeling Research Group, H-1117 Budapest, Hungary}
		\address[t]{Laboratory of Structural Chemistry and Biology, Eötvös University, H-1117, Budapest, Hungary}

\begin{document}

\begin{abstract}
Hexapeptides are increasingly applied as model systems for studying the amyloidogenecity properties of oligo- and polypeptides. It is possible to construct 64 million different hexapeptides from the twenty proteinogenic amino acid residues. Today's experimental amyloid databases contain only a fraction of these annotated hexapeptides.  For labeling all the possible hexapeptides as "amyloidogenic" or "non-amyloidogenic" there exist several computational predictors with good accuracies. It may be of interest to define and study a simple graph structure on the 64 million hexapeptides as nodes when two hexapeptides are connected by an edge if they differ by only a single residue. For example, in this graph, HIKKLM is connected to AIKKLM, or HIKKNM, or HIKKLC, but it is not connected with an edge to VVKKLM or HIKNPM.
In the present contribution, we consider our previously published artificial intelligence-based tool, the Budapest Amyloid Predictor (BAP for short), and demonstrate a spectacular property of this predictor in the graph defined above. We show that for any two hexapeptides predicted to be "amyloidogenic" by the BAP predictor, there exists an easily constructible path of length at most 6 that passes through neighboring hexapeptides all predicted to be "amyloidogenic" by BAP. For example, the predicted amyloidogenic ILVWIW and FWLCYL hexapeptides can be connected through the length-6 path ILVWIW-IWVWIW-IWVCIW-IWVCIL-FWVCIL-FWLCIL-FWLCYL in such a way  that the neighbors differ in exactly one residue, and all hexapeptides on the path are predicted to be amyloidogenic by BAP. 
The symmetric statement also holds true for non-amyloidogenic predicted hexapeptides: for any such pair, there exists a path of length at most 6, traversing only predicted non-amyloidogenic hexapeptides. It is noted that the mentioned property of the Budapest Amyloid Predictor \url{https://pitgroup.org/bap} is not proprietary; it is also true for any linear Support Vector Machine (SVM)-based predictors; therefore, for any future improvements of BAP using the linear SVM prediction technique.
\end{abstract}

\date{}
	
\maketitle	
	
\section*{Introduction} 

Amyloids are misfolded proteins with a well-defined parallel and/or antiparallel repeating  $\beta$-sheet structure \cite{Horvath2019,Taricska2020}. Numerous globular proteins can turn into amyloids in certain physical or chemical environments \cite{Taricska2020}. While amyloids are most frequently mentioned in the context of human diseases \cite{Soto2006}, they can also be functional building blocks in healthy human tissues \cite{Maji2009} or can serve as perspective anti-viral agents \cite{Michiels2020}.

In the last decade, hexapeptides have become a popular class of molecules for modeling and studying the protein amyloid formation: these short peptides are simple enough to be studied in a variety of {\it in vitro} and {\it in silico} systems, yet complex enough to show characteristic amyloid formation changes in numerous studies. Because of their applicability as model systems, experimental data have been collected on hundreds of hexapeptides in relation to their amyloidogenic properties. The creators of the Waltz database \cite{Beerten2015, Louros2020} published 1415 hexapeptides, of which 514 were experimentally labeled as "amyloidogenic" and 901 as "non-amyloidogenic". 

By applying the labeled molecules from the Waltz database for training an artificial intelligence tool, our research group has prepared a support vector machine \cite{Cortes1995} (SVM)-based tool for amyloidogenecity-prediction for hexapeptides \cite{Keresztes2020a}. Our tool, called the Budapest Amyloid Predictor (BAP), is publicly available at  \url{https://pitgroup.org/bap}. We have shown in \cite{Keresztes2020a} that the accuracy of the BAP predictor is 84 \% (and the further quality measures are TPR=0.75, TNR=0.9, PPV=0.8, NPV=0.86; (that is, true positive ratio, true negative ratio, positive predictive value, negative predictive value, resp.).  

A recent review of published amyloid-predictors \cite{Santos2020} lists, among others, Zyggregator \cite{Tartaglia2008a}, AGGRESCAN \cite{Conchillo-Sole2007}, netCSSP \cite{Kim2009a},  APPNN \cite{Familia2015}. Our BAP has the same or better accuracy as the predictors listed in \cite{Santos2020}, as it was shown in \cite{Keresztes2020a}.

The BAP predictor is based on a linear Support Vector Machine (SVM) \cite{Cortes1995}. SVM-based predictors have a much more transparent structure than other artificial intelligence predictors, and this transparency leads to very strong applications. Generally, it is difficult to explain the intrinsic "reason" by which a deep neural network predictor makes a decision or to describe those attributes of the input that lead to a given classification by the network. 

The transparent structure of the SVM predictor BAP \cite{Keresztes2020a} was exploited in our work \cite{Keresztes2022}; where we have identified patterns, describing amyloid-forming hexapeptides very succinctly. For example, we have shown that for any substitution with the 20 proteogenic amino-acids for positions denoted by $x$, all the 
patterns CxFLWx, FxFLFx, or xxIVIV are predicted amyloidogenic, and all the patterns PxDxxx, xxKxEx, and xxPQxx are predicted non-amyloidogenic. We note that any pattern with two x's describes $20^2=400$ hexapeptides, and patterns with four x's describe $20^4=160,000$ hexapeptides. In \cite{Keresztes2022} we have described all such patterns, and also amyloidogenic patterns with restricted choices for the positions of $x$, where the residues were allowed to be selected from polar, non-polar or hydrophobic subsets of the 20 proteogenic amino acids. 

We note that the transparent structure of the Support Vector Machines made it possible to identify different patterns in \cite{Keresztes2022,Keresztes2019,Keresztes2022a}.

In the present contribution, we exploit further the transparent structure of the predictor BAP \url{https://pitgroup.org/bap}. Suppose we want to find a path from a hexapeptide $x$ to another hexapeptide $x'$ through different hexapeptides, such that in each step, we can move from one hexapeptide to another with exactly one different residue position.

\medskip
Note on the terminology: When a sequence of reactions is studied, then the ``pathway'' term is used generally. We apply here the graph-theoretical, more abstract ``path'' term, since we work in the present contribution on a graph.
\medskip

For example, we want to find a path from hexapeptide ILVWIW to hexapeptide FWLCYL, through six-tuples, differing in exactly one residue. An obvious path is generated by changing the amino acids one-by-one from left to right, starting from ILVWIW and finishing at FWLCYL as follows:
\medskip

\centerline{ILVWIW-FLVWIW-FWVWIW-FWLWIW-FWLCIW-FWLCYW-FWLCYL}
\medskip

These paths from one-by-one residue-exchanges can be of interest in peptide synthesis design or following a sequence of point mutations of peptides or protein sequences and measuring or modeling the change of their subsequent chemical or biological properties when only one residue is altered in one step.

Analyzing the effects of subsequent point mutations was done in the literature in the past decades. In \cite{Hoffmueller2000}, three different biologically active peptides were transformed into each other by subsequent single amino acid substitutions, and the intermediaries were analyzed for activity. The authors of \cite{Hoffmueller2000} called the paths formed from the subsequent point-mutated peptides "evolutionary transition pathways".

Paths of one-by-one residue exchanges can be interesting, which connect two predicted amyloidogenic hexapeptides and go through amyloidogenic hexapeptides only.

Similarly, we may want to design paths between the non-amyloidogenic hexapeptides A and B, along which only one residue is changed in each step and which goes through non-amyloidogenic intermediaries only.

In the present contribution, we show the following results for the BAP predictor:

\begin{itemize}
\item{-} All predicted amyloidogenic pairs of hexapeptides, $x$ and $x'$ can be connected by one-by-one exchanged residue-paths of length at most 6, such that the whole path contains only predicted amyloidogenic intermediaries. Moreover, the path can be computed easily.

\item{-} All predicted non-amyloidogenic pairs of hexapeptides, $x$ and $x'$ can be connected by one-by-one exchanged residue-paths of length at most 6, such that the whole path contains only predicted non-amyloidogenic intermediaries. Moreover, the path can be computed easily.
\end{itemize}

We also show that the same results hold for other linear-SVM-based predictors, and not only for our BAP predictor described in \cite{Keresztes2020a}.

We remark that in the case of non-SVM based predictors, it may happen that two predicted amyloidogenic sequences cannot be connected by entirely amyloidogenic paths of {\it any} length and the same holds for the non-amyloidogenic case, too. For example, in non-SVM-based predictors, it may happen that all the neighbors of an amyloidogenic peptide A are predicted to be non-amyloidogenic; consequently, A cannot be connected by an entirely amyloidogenic path to any other amyloidogenic peptide.

We also remark that we do not state anything on paths connecting amyloidogenic hexapeptides with non-amyloidogenic ones.

\section*{Methods}

Here, we first formalize our problem setting and solution, and then we will make some remarks on possible generalizations. 

All our definitions and methods or algorithms will be specified for hexapeptide sequences, but they are easily generalizable for shorter or longer amino acid sequences of a given length.

First, we define the mutation-graph $M$ on the hexapeptide sequences:

\begin{definition}
The vertices of the mutation-graph $M$ are the $20^6=64$ million hexapeptides formed from the 20 proteogenic amino acids. The vertices are referred to using their length-6 amino acid sequences with the one-letter codes. Two vertices of $M$ are connected by an edge if they differ in exactly one amino acid in the same position.
\end{definition}

\begin{example}
Node ILVWIW is connected by an edge to ALVWIW, or to IAVWIW, or to ILVWID, but not to IDDWIW. 
\end{example}

We note that paths in this graph $M$ were called "evolutionary transition pathways" in \cite{Hoffmueller2000}. We simply call them "paths'' in $M$. The length of a path is the number of edges in it.

It is easy to see that in each position, we can make 19 different substitutions (the original amino acid can be substituted by any of the remaining 20-1=19 proteogenic amino acids), and since we have six positions, every vertex is connected to 6 $\times$ 19 = 114 other nodes, which represent exactly 114 hexapeptides.

Next, we partition the vertices of $M$ in two classes: amyloidogenic and non-amyloidogenic. That is, each vertex is an element of one and only one of those classes. The partitioning is done by the Budapest Amyloid Predictor, described in details in \cite{Keresztes2020a}.

\subsection*{The Budapest Amyloid Predictor and the Amyloid Effect Matrix}

Here, we succinctly describe the BAP predictor with details needed to prove our statement and to show our method for finding the paths, leading entirely in one of the two partition classes. The details of the construction of the Budapest Amyloid Predictor, the evaluation of its correctness, and the comparison with other predictors were described in detail in \cite{Keresztes2020a}.

The BAP predictor uses a linear Support Vector Machine (SVM) \cite{Cortes1995} for decisions. A linear SVM computes the sign of the value
$$\sum_{i=1}^nw_iz_i+b\eqno{(1)}$$
and it makes a decision based on this sign. Here, the coefficients $w_1,w_2,\ldots,w_n$ and $b$ are real numbers computed from the training data, and $z_1,z_2,\ldots,z_n$ represent the input values. For example, if for a given input $z=(z_1,z_2,\ldots,z_n)$ the value of (1) is non-negative, the SVM outputs ``yes'' otherwise ``no''.  

The Budapest Amyloid Predictor \cite{Keresztes2020a} (available at \url{https://pitgroup.org/bap} applied the Waltz dataset \cite{Beerten2015, Louros2020} for training and testing an SVM, where each of the 20 proteogenic amino acids was represented as a (highly redundant) length-553 vector $Z$, corresponding to 553 properties of AAindex \cite{Kawashima2008}. Therefore, a hexapeptide was represented by six concatenated $Z$ vectors; their combined length is $6 \times 553 = 3318=n$. 

With  $\ell=553$, equation (1) can be written as 

$$
\sum_{i=1}^{6\ell}w_iz_i+b=\sum_{j=1}^6  \ \ \sum_{i=(j-1)\ell+1}^{j\ell}w_iz_i+b \eqno{(2)}
$$

If the value of (2) is negative (i.e., its sign is -1), the hexapeptide is predicted to be non-amyloidogenic if it is positive or 0, it (its sign is 1 or 0) is predicted to be amyloidogenic. 

Here, index $j$ refers to amino acid $j$ in the hexapeptide For $j=1,2,\ldots,6$. Since the $\ell=553$ $z_i'$s are determined by the $j^{th}$ amino acid of the hexapeptide, and this way, all the possible $6 x 20=120$ second sums in (2) (for six positions and 20 amino acids) can be pre-computed. 

 Table 1 lists these pre-computed values: the 6 values of $j$ correspond to the columns, the amino acids to the rows. In other words, Table 1, which is called the ``Amyloid Effect Matrix'' in \cite{Keresztes2020a}, describes the position-depending contributions of amino acids to the value of (2).

\begin{figure}[H]
	\begin{center}
		\includegraphics[width=8cm]{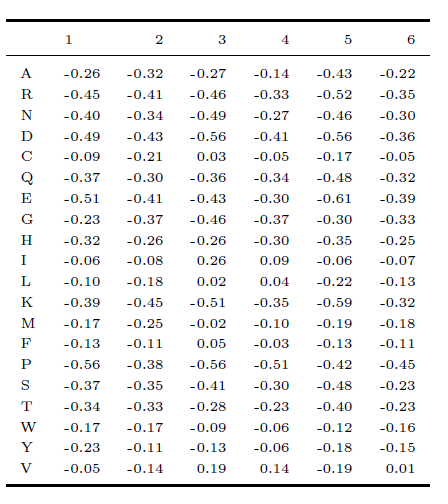}
		{\\
Table 1: The Amyloid Effect Matrix \cite{Keresztes2020a}. The pre-computed values from equation (2) are listed in the rows corresponding to the amino acids. The columns are  corresponded to the positions in the hexapeptide. }
	\end{center}
\end{figure}

Table 1 facilitates the easy ``by hand'' computation of sum (2) and making a decision on its amyloidogenecity. For example, if we want to make a prediction on YVSTSY, then we need to take the value from column 1, corresponding to Y (i.e., $-0.23$, and from column 2, corresponding to V (-0.14), from column 3, corresponding to S (-0.41), from column 4 in row of T (-0.23), from column 5 in the row of S (-0.48), and from column 6 corresponding to Y (-0.15), add them up, and add $b=1.083$ to the sum: $-0.23-0.14-0.41-0.23-0.48-0.15+1.083=-0.557$; therefore, YVSTSY is predicted to be non-amyloidogenic. 

One can simply order the amino acids in each position of the hexapeptides according to their contribution to sum (2), as in Table 2.

\begin{figure}[H]
	\begin{center}
		\includegraphics[width=9cm]{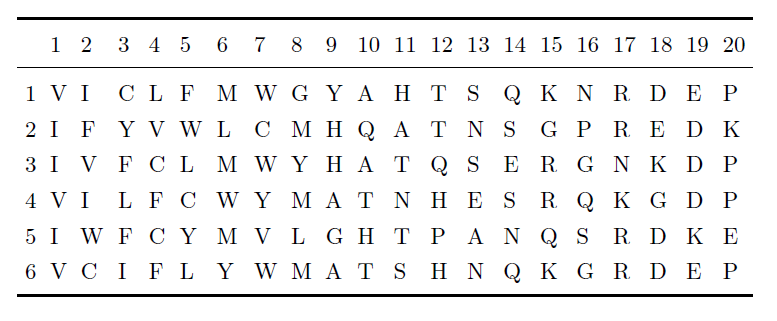}
		{\\ Table 2: The position-specific amyloidogenecity order of the amino acids, decreasing from left to right, \cite{Keresztes2020a}. }
	\end{center}
\end{figure}

Table 2 has some very practical applications for amyloidogenecity prediction. If we have one predicted amyloidogenic hexapeptide $x$, then we can easily make numerous other predicted amyloidogenic hexapeptides from $x$, simply by replacing any amino acid in a given position by one, which is situated left to the original one in its row in Table 2. More exactly, if hexapeptide $x$ is predicted to be amyloidogenic, and its 3rd amino acid is $Y$, then $Y$ can be exchanged to either of I, V, F, C, L, M, or W, the resulting hexapeptide $x'$ will always be predicted to be amyloidogenic. This is true since Table 2 contains the orderings of the amino acids in each position according to their contribution in Table 1, and if we exchange $Y$ to anything from its left in row 3 of Table 2, then we increase the value of the sum of (2), relative to that of $x$. Since the value of the sum in the case of $x$ was positive, its increased value will also be positive, i.e., the decision of the SVM will be ``amyloidogenic''.

Similarly, in the case of a hexapeptide $n$, predicted to be non-amyloidogenic, if we exchange any amino acids located to the right from the original in Table 2, then the new prediction will also be non-amyloidogenic. For example, if the 4th amino acid of $n$ is $E$, and we exchange $E$ to any of the S, R, Q, K, G, D, P, then the value of sum (2) will be decreased, and, consequently, the prediction will remain non-amyloidogenic.

The description of the path constructions will be more convenient by introducing two simple operators, for $i=1,2,3,4,5,6$, and for any two (not necessarily distinct) amino acids $X$ and $X'$:

\medskip
$${\rm MAX}_i(X,X')= \hbox{ In row i of Table 2 the leftmost one of } (X,X')$$
\medskip
$${\rm MIN}_i(X,X')= \hbox{ In row i of Table 2 the rightmost one of } (X,X').$$
 \medskip
If $X=X'$ in any of the two operators, then the output value is $X=X'$.

The MAX and MIN terms refer to the amyloidogenecity of the amino acids in position $i$.
 \medskip
 
 Let us call the value (2) of a hexapeptide $x$ its amyloidogenecity value, and let us denote it by $A(x)$. If $A(x)\geq 0$, then $x$ is predicted to be amyloidogenic, otherwise non-amyloidogenic.

\section*{Results}

Here, we show how to connect any two hexapeptides of the same amyloidogenecity prediction with a path of length at most 6, going on the same class as their endpoints in graph $M$, the mutation graph.

\subsection*{Constructing paths through the amyloidogenic hexapeptides}

 Suppose we have two hexapeptides, $x=(X_1,X_2,X_3,X_4,X_5,X_6)$ and $x'=(X'_1,X'_2,X'_3,X'_4,X'_5,X'_6)$, both predicted to be amyloidogenic by BAP.
 For simplicity, we will call the $X_i$ amino acids ``coordinates'' of $x$.
 
 Now we show that there exists an easily constructible path of length at most 6 in graph $M$, such that all vertices of the path are predicted amyloidogenic.
 
 \medskip

{\noindent \bf Case 1} (the easy case): Suppose that $x'$ is ``coordinate-wise more amyloidogenic'' than $x$ in the following sense: for all  $i=1,2,3,4,5,6$, either  $X_i=X'_i$, or $X'_i$ is situated left from $X_i$ in row $i$ of Table 2; that is, $X_i$ is less amyloidogenic in position $i$ than $X'_i$. Then, if we change $X_i$ to $X'_i$ in position $i$, for $i=1,2,3,4,5,6$, then we go through a path from $x$ to $x'$ in graph $M$ such that the value of $A$ on the nodes, describing the amyloidogenecity will be monotone increasing. Therefore, all nodes on the path will be predicted to be amyloidogenic. Note that the length of this path is at most 6: when the same amino acid appears in the same coordinates, no change is needed; when every coordinate is different, then 6 changes are needed.

Formally:
$$
	A(x)=A(X_1,X_2,X_3,X_4,X_5,X_6)\leq A(X'_1,X_2,X_3,X_4,X_5,X_6)\leq$$
	$$\leq A(X'_1,X'_2,X_3,X_4,X_5,X_6)\leq\ldots\leq A(X'_1,X'_2,X'_3,X'_4,X'_5,X'_6)=A(x')$$

{\noindent \bf Case 2} (the general case): When the assumptions of Case 1 are not satisfied, we reduce the problem to two applications of path-finding in Case 1.

Our strategy is as follows: 

\begin{itemize}
	\item[Step I:] First, we connect node  $x=(X_1,X_2,X_3,X_4,X_5,X_6)$ to node  $$x_{MAX}=({\rm MAX}_1(X_1,X'_1),{\rm MAX}_2(X_2,X'_2),{\rm MAX}_3(X_3,X'_3),$$
	$${\rm MAX}_4(X_4,X'_4),{\rm MAX}_5(X_5,X'_5),{\rm MAX}_6(X_6,X'_6)),$$ exactly as in Case 1, since they satisfy the assumptions.
	
	\item[Step II:] Second, we connect $x'$ to $x_{MAX}$, as in Case 1, since they satisfy the assumptions.
\end{itemize}

Now, we detail that in both Step I and Step II, the requirements of Case 1 are satisfied. Since both $x$ and $x'$ are amyloidogenic, and since both $A(x)\leq A(x_{MAX})$ and $A(x')\leq A(x_{MAX})$ hold, $x_{MAX}$ is also ``coordinate-wise more amyloidogenic'' than both $x$ and $x'$. 

In other words, because of the definition of the ${\rm MAX}_i$ operators, the coordinates of $x_{MAX}$ are left in Table 2 from the coordinates $x$ and $x'$ in each row. Therefore, in Step I, the procedure of Case 1 can be applied for connecting $x$ to $x_{MAX}$;  and in Step II, the procedure of Case 1 can be applied to connect $x'$ to $x_{MAX}$. Since the paths are undirected, we take the path $x$ to $x_{MAX}$ and further  to $x'$.

Now we show that the combined length of the path from $x$ to $x_{MAX}$, and from $x_{MAX}$ to $x'$ is at most 6: It is easy to verify that for all $i$: ,${\rm MAX}_i(X_i,X'_i)$ is either equal to $X_i$ or $X'_i$, so if an exchange is needed in Step I in coordinate $i$, then no change is needed in Step II in coordinate $i$, and a symmetric remark is also true for Step II and Step I.  

\begin{example}
	
	Let us connect hexapeptides $x=$CVFFFF to $x'=$LYCLCI by a predicted amyloidogenic path. Both $x$ and $x'$ are predicted amyloidogenic. Case 1 cannot be applied (one can see it easily from Table 2), so we need $x_{MAX}=$CYFLFI. So, we first connect $x$ to $x_{MAX}$:	
	$$CVFFFF-CYFFFF-CYFLFF-CYFLFI$$
	Then $x'$ to $x_{MAX}$:
	$$LYCLCI-CYCLCI-CYFLCI-CYFLFI$$
	The full path is:
	$$CVFFFF-CYFFFF-CYFLFF-CYFLFI-CYFLCI-CYCLCI-LYCLCI$$
	
\end{example}

\subsection*{Constructing paths through the non-amyloidogenic hexapeptides}

The proof of this case is the repetition of the construction above, with the obvious changes. For completeness, we give here the proof.

Suppose we have two hexapeptides, $x=(X_1,X_2,X_3,X_4,X_5,X_6)$ and $x'=(X'_1,X'_2,X'_3,X'_4,X'_5,X'_6)$, both predicted to be non-amyloidogenic by BAP.

We show that there exists an easily constructible path of length at most 6 in graph $M$, such that all vertices of the path are predicted non-amyloidogenic.

\medskip
{\noindent \bf Case 1} (the easy case): Suppose that for all  $i=1,2,3,4,5,6$, either  $X_i=X'_i$, or $X'_i$ is situated right from $X_i$ in row $i$ of Table 2; that is, $X'_i$ is less amyloidogenic in position $i$ than $X_i$. Then if we change $X_i$ to $X'_i$ in position $i$, for $i=1,2,3,4,5,6$, then we go through a path in graph $M$ from $x$ to $x'$  such that the value of $A$ on the nodes, describing the amyloidogenecity, will be monotone decreasing.  Note that the length of this path is at most 6: when the same amino acid appears in the same coordinates, no change is needed; when every coordinate is different, then 6 changes are needed.

More formally:

$$
A(x)=A(X_1,X_2,X_3,X_4,X_5,X_6)\geq A(X'_1,X_2,X_3,X_4,X_5,X_6)\geq$$
$$\geq A(X'_1,X'_2,X_3,X_4,X_5,X_6)\geq\ldots\geq A(X'_1,X'_2,X'_3,X'_4,X'_5,X'_6)=A(x')$$

{\noindent \bf Case 2} (the general case): When the assumptions of Case 1 are not satisfied, we reduce the problem to two applications of path-finding in Case 1.

Our strategy is as follows: 

\begin{itemize}
	\item[Step I:] First, we connect node  $x=(X_1,X_2,X_3,X_4,X_5,X_6)$ to node  $$x_{MIN}=({\rm MIN}_1(X_1,X'_1),{\rm MIN}_2(X_2,X'_2),{\rm MIN}_3(X_3,X'_3),$$
	$${\rm MIN}_4(X_4,X'_4),{\rm MIN}_5(X_5,X'_5),{\rm MIN}_6(X_6,X'_6)),$$ exactly as in Case 1, since they satisfy the assumptions.
	
	\item[Step II:] Second, we connect $x'$ to $x_{MIN}$, as in Case 1, since they satisfy the assumptions.
\end{itemize}

Now, we detail that in both Step I and Step II, the requirements of Case 1 are satisfied. Since both $x$ and $x'$ are amyloidogenic, and since both $A(x)\geq A(x_{MIN})$ and $A(x')\geq A(x_{MIN})$ hold, $x_{MIN}$ is also coordinate-wise less amyloidogenic than both $x$ and $x'$. 

In other words, because of the definition of the ${\rm MIN}_i$ operators, the coordinates of $x_{MIN}$ are right in Table 2 from the coordinates $x$ and $x'$ in each row. Therefore, in Step I, the procedure of Case 1 can be applied for connecting $x$ to $x_{MIN}$;  and in Step II, the procedure of Case 1 can be applied to connect $x'$ to $x_{MIN}$. Since the paths are undirected, we take the path $x$ to $x_{MIN}$ and further  to $x'$.

Now we show that the combined length of the path from $x$ to $x_{MIN}$, and from $x_{MIN}$ to $x'$ is at most 6: It is easy to verify that for all $i$: ,${\rm MIN}_i(X_i,X'_i)$ is either equal to $X_i$ or $X'_i$, so if an exchange is needed in Step I in coordinate $i$, then no change is needed in Step II in coordinate $i$, and a symmetric remark is also true for Step II and Step I.

\section*{Conclusions} 

We have shown that the linear SVM predictors for peptides have a very transparent structure that can be used to design mutational pathways within the predicted classes. More specifically, we have used the Budapest Amyloid Predictor \cite{Keresztes2020a} to partition 64 million possible hexapeptides into two classes: predicted amyloidogenic and predicted non-amyloidogenic, and we have shown that any two members of each class can be connected by a mutation pathway of length at most 6 that lies entirely within the same class, i.e., amyloidogenic or non-amyloidogenic. For the construction, we used Table 2, defined by the Budapest Amyloid Predictor. The exact same result can be obtained using any other updated version of Table 2, so our results here are not specific to the Budapest Amyloid Predictor.  
	
\section*{Data availability} All data are included in the text. 

\section*{Funding}

VG was partially funded by the Ministry of Innovation and Technology of Hungary from the National Research, Development and Innovation Fund, financed under the  ELTE TKP 2021-NKTA-62 funding scheme.

\section*{Author Contribution} LK, ES, AP, VF and VG have initiated the study and evaluated results, LK and ES constructed the SVM for the prediction, BV has constructed the webserver, VG has overseen the work and wrote the first version of the paper; all authors have reviewed the article. AP, VF and VG secured funding.

\section*{Conflicting interest} The authors declare no conflicting interests.

%\bibliography{v:/vince/CIKKEK/medl}
%\bibliographystyle{unsrtnat}

\eject

\end{document}